\documentclass[]{tMOP2e}

\usepackage{color}

\begin{document}
%

\markboth{M.R. Bionta {\it et al.}}{Journal of Modern Optics}


\title{Laser-induced electron emission from a tungsten nanotip: identifying above threshold photoemission using energy-resolved laser power dependencies}


\author{M.R. Bionta$^{a}$, B. Chalopin$^{a}$$^{\ast}$, J.P. Champeaux$^{a}$, S. Faure$^{a}$, A. Masseboeuf$^{b}$, P. Moretto-Capelle$^{a}$, and B. Chatel$^{a}$\thanks{$^\ast$Corresponding author. Email: benoit.chalopin@irsamc.ups-tlse.fr\vspace{6pt}}
\\\vspace{6pt}  
$^{a}${\em{LCAR-UMR 5589-Universit\'e Paul Sabatier Toulouse III-CNRS, 118 Route de Narbonne, Bat 3R1B4, 31062 Toulouse Cedex 9, France}};
$^{b}${\em{CEMES-UPR 8011-CNRS, 29 rue Jeanne Marvig, BP 94347, Toulouse, Cedex 4, France}}\\\vspace{6pt}\received{v3.4 released October 2010} }

\maketitle

\begin{abstract}
We present an experiment studying the interaction of a strongly focused 25~fs laser pulse with a tungsten nanotip, investigating the different regimes of laser-induced electron emission. We study the dependence of the electron yield with respect to the static electric field applied to the tip. Photoelectron spectra are recorded using a retarding field spectrometer and peaks separated by the photon energy are observed with a 45~\% contrast. They are a clear signature of above threshold photoemission (ATP), and are confirmed by extensive spectrally resolved studies of the laser power dependence. Understanding these mechanisms opens the route to control experiment in the strong-field regime on nanoscale objects.
\bigskip

\end{abstract}

\bigskip

\section{Introduction}
The development of coherent control schemes through the manipulation of atomic and molecular quantum dynamics \cite{rabitz_whither_2000,meshulach_coherent_1998} both theoretically and experimentally, has been a very active field of research in the last two decades. This has led to the availability of robust and selective methods of performing population transfer in quantum systems through the control of laser-matter interactions. In order to fully control these systems, great effort has been done to increase the number of available parameters, leading to important technological breakthroughs \cite{monmayrant_newcomers_2010} that has opened the possibility to design arbitrarily shaped optical waveforms in the fs-ps time domain.

In parallel, the development of nanostructure devices provides the opportunity to add new parameters of control, including light confinement, which facilitates the capacity to reach the strong-field domain and the possibility to work on single objects. 
Strong field phenomena have been studied in solid-state nanostructures such as ponderomotive acceleration and carrier-envelope phase effects in photoemission from metallic surface \cite{kupersztych_ponderomotive_2001}, dielectric nanospheres \cite{zherebtsov_controlled_2011}, or gold nanoparticles \cite{PhysRevLett.108.237602}.

Sharp metallic tips are emerging as a test bed for exploring various strong-field phenomena \cite{Kruger2012} revealing the multiphoton photoemission (MPP) and above threshold photoemission (ATP) \cite{barwick_laser-induced_2007, ropers_localized_2007,schenk_strong-field_2010,kruger_attosecond_2011, park_strong_2012}, and optical field emission \cite{herink_field-driven_2012} regimes. Strong-field photoemission from such metallic nanostructrures benefit from the optical field enhancement around sharp objects, which reduces the necessary laser intensity needed to reach this regime, compared to the case of atomic and molecular systems. Specifically, the optical field enhancement at the apex reduces the laser intensities needed to reach the strong-field regime. Additionally, the combination of a DC field with the AC field provided by the laser pulse itself adds a supplemental control parameter, as the DC field will modify the behavior of the potential barrier. 

In this paper, we revisit the multiphoton regime of the interaction of an ultrashort laser pulse with a tungsten nanotip, by carefully studying the photoelectron energy-resolved power law.

We first recall the different regimes of photoemission, then describe our experimental details; we present above threshold photoemission photoelectron spectra, then discuss of energy resolved laser power dependencies.

\section{Different Regimes of Laser Induced Emission}

\begin{figure}
\begin{center}
\begin{minipage}{100mm}
\begin{center}
\includegraphics[width = 50mm]{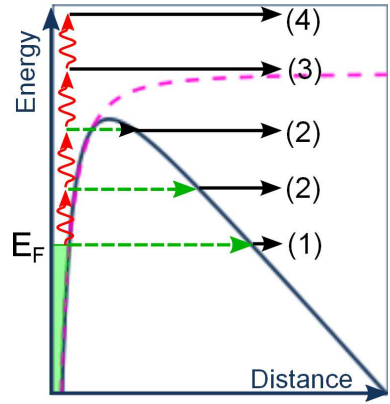}%
\caption{\label{principles}(color online) Principle of different regimes of emission from a metallic surface with a static field and a laser field of moderate intensity. The pink dashed line represent field-free potential. The solid blue line is the potential modified by the static field. $E_{F}$ is the Fermi level of electrons in the metal. Path (1):  Static field emission: an electron simply tunnels through the potential barrier. Path (2): Photo field emission: the system absorbs 1 or 2 photons, increasing the electron's probability to tunnel through the potential barrier. Path (3): The system absorbs 3 photons and the electron can overcome the potential barrier. Path (4): Above threshold photoemission (ATP): the system absorbs more photons than necessary for the electron to overcome the potential barrier.}
\end{center}
\end{minipage}
\end{center}
\end{figure}

Static field emission from a nanotip is described accurately by the Fowler-Nordheim theory \cite{Fowler-nordheim}, which calculates the tunneling probability (path 1 on Figure \ref{principles}) and predicts the electron yield as a function of the static electric field $E_{DC}$. By measuring the electron current as a function of the applied voltage (called $V_{tip}$ for tip bias), we can use this theory to retrieve the ratio $\beta$ between the static electric field $E_{DC}$ and $V_{tip}$, with $E_{DC} = \beta V_{tip}$. $\beta$ is related to the tip radius $r$ as $\beta = \frac{1}{kr}$ with $k$ being a dimensionless factor depending on the shape of the tip and the extraction geometry \cite{gomer_field_1961}, with typical values ranging from 1 to 20.

The photoemission regimes can be described in the framework of Keldysh theory \cite{keldysh_ionization_1965} using the characteristic parameter $\gamma$ that separates two limiting regimes, the multiphoton regime ($\gamma\gg1$) and the tunneling regime ($\gamma\ll1$). The latter is termed optical field emission for metals. For a metal with a work function $\phi$, the Keldysh parameter, $\gamma$, is given by $\gamma=\sqrt{\frac{\phi}{2U_{p}}}$, where $U_{p}$ is the ponderomotive energy and $U_{p}\propto\frac{I}{\omega^{2}}$ and corresponds to the mean kinetic energy of a free electron oscillating in a monochromatic light field of central frequency $\omega$ and peak intensity $I$. In all the experiments described below, the laser intensity is moderate and $\gamma>1$. Therefore, the optical field emission regime can be excluded. 

Combining a strong DC field with a weak laser intensity can first lead to photofield emission. In this regime, studied extensively\cite{hommelhoff_ultrafast_2006,Yanagisawa2011}, an electron absorbs one or two photons, and tunnels through the potential barrier (paths 2 on Figure \ref{principles}).


For a small DC field and sufficient laser intensity, electrons can be emitted by the absorption of enough photons to overcome the potential barrier, $n\hbar\omega>\Phi$, where $n$ is the number of absorbed photons (path 3 in Figure~\ref{principles}). If the laser intensity is strong enough, more photons than necessary can be absorbed, leading to electrons of higher energies (path 4 in Figure~\ref{principles}). This phenomenon is called above threshold ionization in atoms and molecules and has been extensively studied in the last two decades \cite{agostini_free-free_1979,PhysRevLett.59.1092,Mainfray1991,blaga_strong-field_2009}, while its analog in solid-state physics, above threshold photoemission (ATP), has only been recently demonstrated \cite{banfi_experimental_2005}. Field enhancement around nanostructures allows this regime to be reached before the damage threshold of the material. Both experimental \cite{schenk_strong-field_2010} and theoretical \cite{Wachter_Burgdorfer_PRB2012} studies on ATP have been performed for the case of tungsten metallic tips.

\section{Experimental Details}

\begin{figure}
\begin{center}
\begin{minipage}{150mm}
\subfigure[]{
\resizebox*{75 mm}{!}{\includegraphics{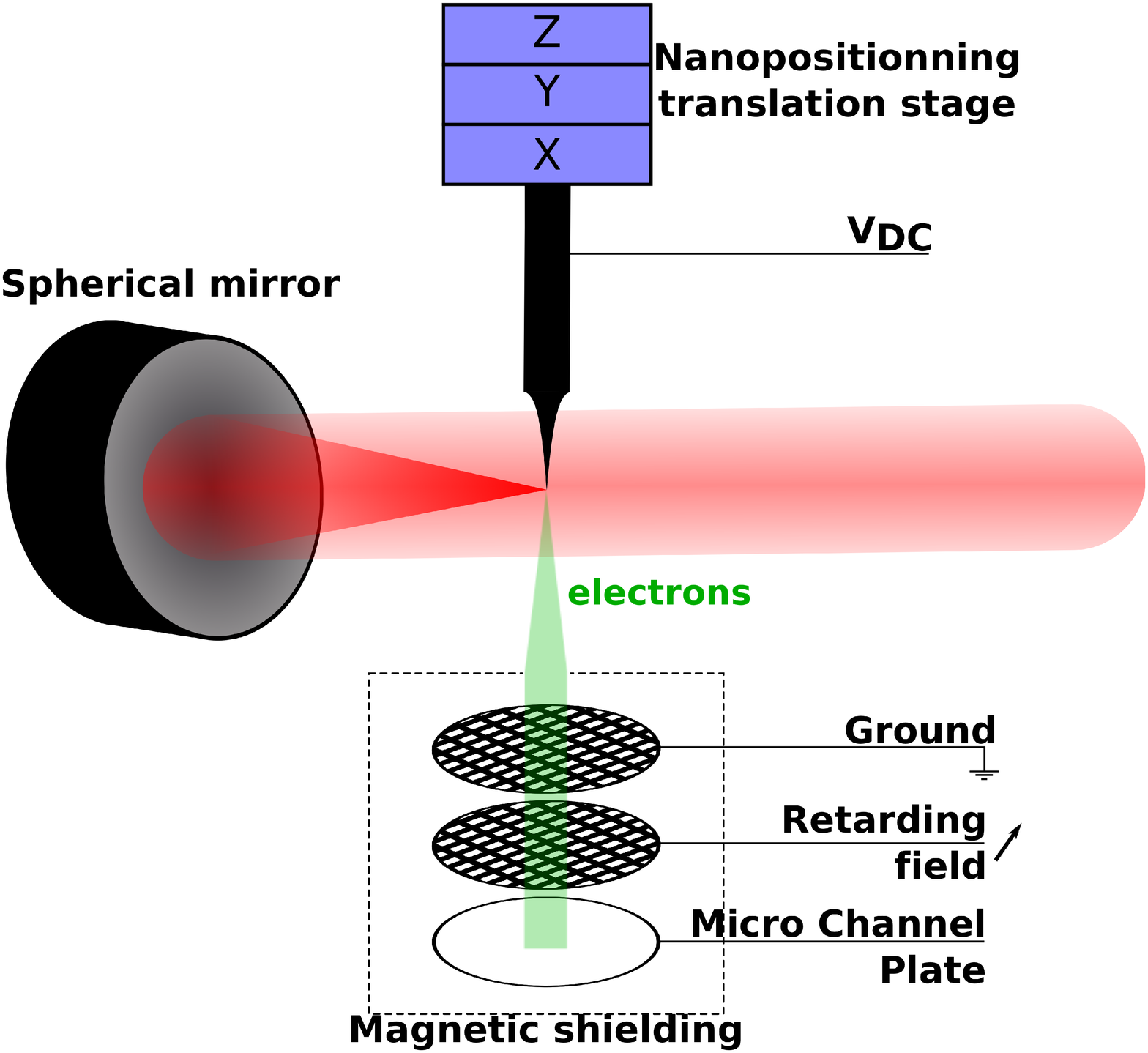}}}%
\subfigure[]{
\resizebox*{75 mm}{!}{\includegraphics{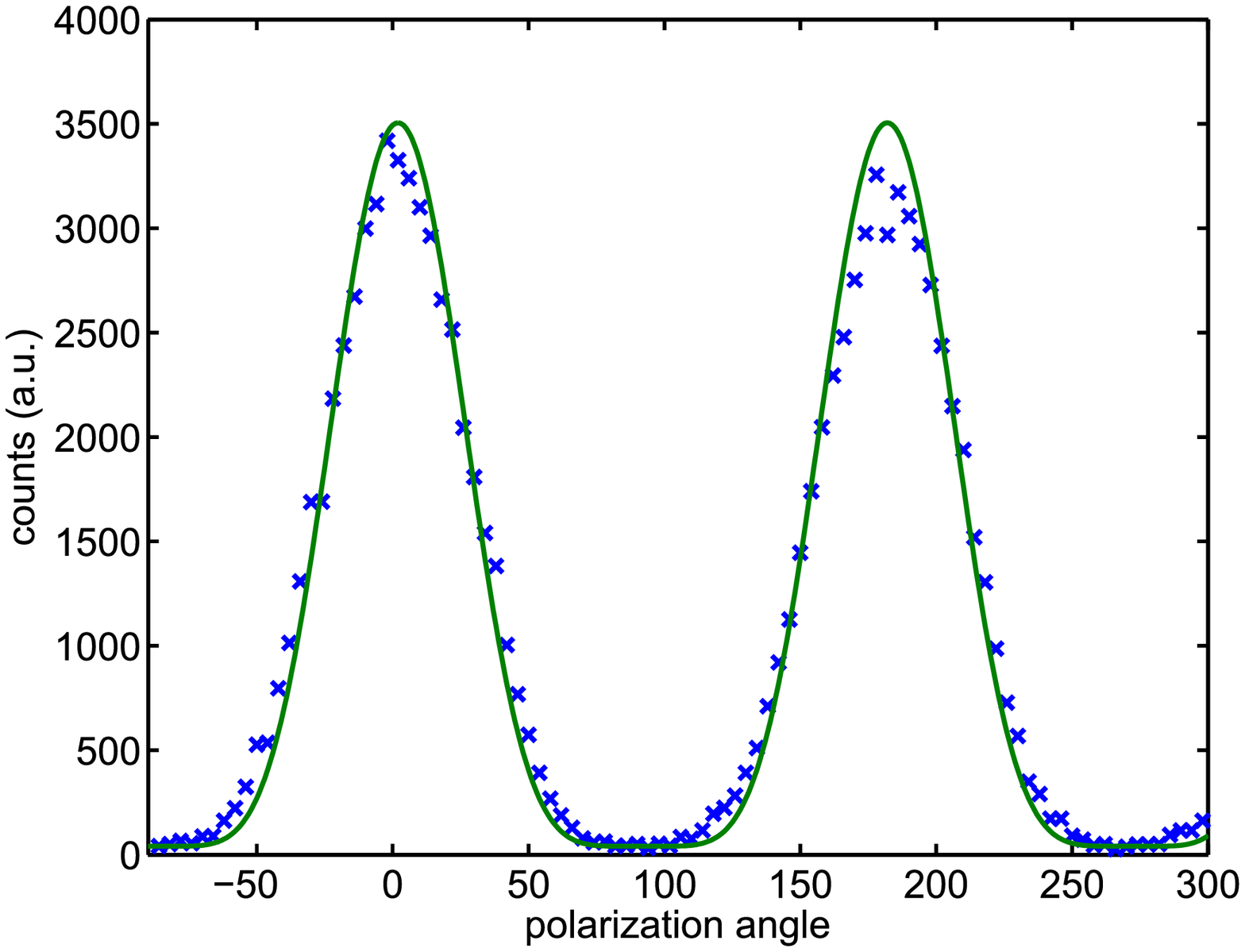}}}%
\begin{center}
\begin{minipage}{100mm}
\caption{\label{fig:setup}(color online) (a) Experimental set up: A femtosecond laser (62~MHz repetition rate, centered at 800 nm, with $\sim$25~fs pulse duration) is tightly focused onto the nanotip to induce electron emission. Nanopositioning translation stages allow precise placement of the tip into the focus. Electrons are detected using a field retarding spectrometer. Laser intensity and polarization are carefully controlled upstream. (b) Electron yield versus relative polarization of the laser with the tip axis. When properly aligned, polarizations parallel to the tip axis (polarization angles of 0 and 180) have the highest electron yield, and polarizations perpendicular to the tip axis (polarization angles of 90 and 270) have negligible emission. The blue x's are experimental data. The green line is a fit.}%
\end{minipage}
\end{center}
\end{minipage}
\end{center}
\end{figure}

Polycrystalline tungsten nanotips are formed using a standard electrochemical etching process in KOH with a radius on the order of $\sim$50-100~nm. The radius of the tip is confirmed both via scanning electron microscope imaging techniques as well as using a Fowler-Nordheim fit from static electron emission, from which the $\beta$ parameter described above is retrieved. The tips are cleaned by either sending a pulse of current ($\sim$3.5 A and $\sim$1 s) through the tip (flashing) or by laser heating (several tens of mW of laser light shown on the tip for several minutes). This leads to the evaporation of the surface contamination of the tip.

A sketch of the experimental details is depicted in Figure~\ref{fig:setup}(a). The direct output of a Ti:sapphire oscillator with 800~nm center wavelength (1.55~eV photon energy), $\sim$25~fs duration and 62~MHz repetition-rate, is focused onto the apex of the tungsten nanotip. The chirp of the laser pulse is carefully controlled with chirped mirror pairs to compensate for dispersion induced by the system and measured with a spectral phase interferometry for direct electric-field reconstruction (SPIDER) system \cite{iaconis_spectral_1998}. The laser intensity is controlled with a half-wave plate polarizer energy throttle; the polarization is cleaned by a second polarizer and then regulated by another half-wave plate. The laser beam is focused with an on-axis silver-coated spherical mirror with a focal length of 4.5~mm to a beam waist of $\sim$2~$\mu$m corresponding to a power density on the order of 1 to 20$\times10^{11}$~W/cm$^{2}$. This corresponds to a $U_{p}$ of 6 to 120~meV. The Keldysh parameter, $\gamma$, is therefore between 4 and 20. without taking into account the optical field enhancement. This value is difficult to measure. In previous experiments \cite{Hommelhoff2006,Kruger2012}, Hommelhoff and coworkers indirectly evaluated the value of the optical field enhancement and found it to be between 3 and 5. Assuming a value of 4 for the field enhancement, the Keldysh parameter would be between 1 and 5 in our experiment.

The experiment is performed in a stainless steel ultra-high vacuum chamber with a pressure on the order of 10$^{-10}$~mbar. A metal-plate with a 1~mm pinhole is placed $\sim$7~mm from the apex of the tip to define a ground potential and an additional small DC field--between 0 and 600~V--is applied to the tip (hereafter referred to as $V_{tip}$). A field retarding spectrometer based on a mesh grid with an adjustable voltage combined with a double stage Micro Channel Plate (MCP) measures the kinetic energy distribution of the electrons with an estimated resolution of dE/E~$\sim5\times10^{-3}$. The spectrometer is placed in a magnetic shield to isolate low-energy electrons from external magnetic fields. It allows either to detetct all electrons (integrated measurement), or to perform energy selective measurement when the retarding voltage is scanned.

Alignment of the tip into the laser focus is achieved using nanopositioning translation stages (from Attocube systems AG) in three dimensions each with nanometer resolution and a range of a few millimeters. The alignment is checked using the diffraction seen by a long-distance microscope objective on a CCD camera mounted perpendicular to laser propagation direction. Confirmation of the exact alignment is corroborated by rotating the polarization of the laser light. As stated by Barwick {\it et al.} \cite{barwick_laser-induced_2007}, when correctly aligned, electron emission is negligible for polarizations perpendicular to the tip axis. Figure~\ref{fig:setup}(b) shows electron yield versus $\theta$, the laser polarization angle with respect to the tip axis. The laser power was 150~mW (8$\times10^{11}$~W/cm$^2$ peak intensity) and $V_{tip}$ is 50~V. The blue x's represent experimental data, and the green line is a fit of the form $\cos^{2k}(\theta)$, where $k=2.8$ is the power exponent measured independently as explained in section~\ref{sec:results}. As expected, polarizations parallel to the tip axis (polarization angles of 0 and 180) have the highest electron yield, and polarizations perpendicular to the tip axis (polarization angles of 90 and 270) have negligible emission. Maximizing the contrast of this curve leads to proper alignment. When misaligned, a secondary peak at perpendicular polarization arises; when the tip is damaged, the contrast decreases.

\section{Results\label{sec:results}}

The first set of experiments investigates the influence of laser intensity on electron emission for various experimental parameters.

\begin{figure}
\begin{center}
\begin{minipage}{100mm}
\includegraphics[width = 100 mm]{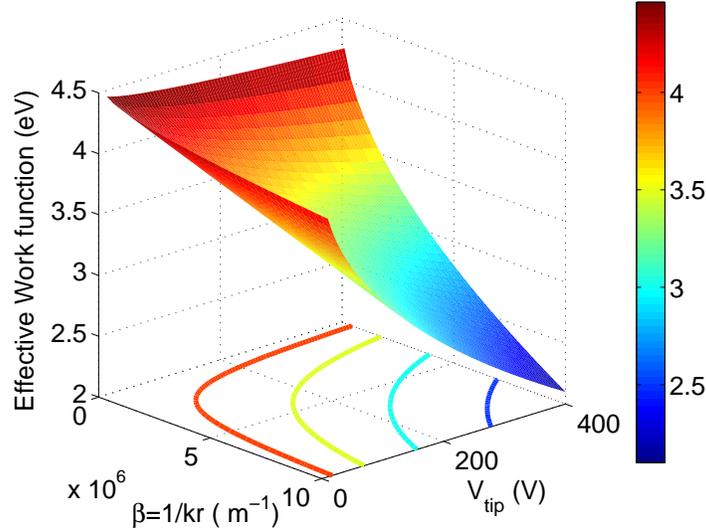}%
\caption{\label{fig:schottky} (color online) The effective work function $\Phi_{eff}$ of the polycrystalline tungsten tip as a function of the applied tip bias ($V_{tip}$) and the enhancement factor $\beta = 1/kr$, due to the sharpness of the object. The presence of a static electric field decreases the barrier height (Schottky effect).}
\end{minipage}
\end{center}
\end{figure}
In the case of a polycrystalline tungsten tip, the work function is about 4.5~eV \cite{gomer_field_1961} and is slightly dependent of the surface absorption during the experiment. Working in ultra-high vacuum as well as cleaning the tip before the experiment greatly reduces this effect. Since the energy of one photon is 1.55~eV, without the DC field, an electron must absorb at least 3 photons to overcome the work function without tunneling. Applying a DC field $E_{DC}= \beta V_{tip}$ affects the yield of laser-induced field-emission from the tip in different ways: 
emitted electrons are accelerated by $V_{tip}$ up to a mean kinetic energy of $e \cdot V_{tip}$ where $e$ is the elementary charge, but $V_{tip}$ also affects the dependence of yield on laser intensity. At least two effects can be taken into account: the presence of the Schottky barrier and the possibility to favor tunneling effect.

   First, the image potential (also known as the Schottky barrier) decreases the work function by a value equal to $\sqrt{\frac{e^{3}\beta V_{tip}}{4\pi\varepsilon_0}}$ where $\beta$ the enhancement factor and $e$ is the electron charge while $\varepsilon_0$ is the vacuum permittivity. The enhancement factor is strongly dependent on the tip geometry and can reach values varying from 10$^{6}$ to 10$^{7}$~m$^{-1}$ from tip to tip. The effective work function is therefore $\Phi_{eff} = \Phi - \sqrt{\frac{e^{3}\beta V_{tip}}{4\pi\varepsilon_0}}$. Figure \ref{fig:schottky} plots $\Phi_{eff}$ as a function of both $V_{tip}$ and $\beta$. A reduction of the work function on the order of one photon energy can be reached via field enhancement near a very sharp object, which strongly changes the electronic response of the system. In our experiment, $\beta$ is typically in the order of $2\times10^{6}~m^{-1}$ and $V_{tip}$ is on the order of 100~V or less, which gives $E_{DC}$ on the order of 10$^8$~V/m. This leads to a work function reduced by 0.5~eV.  
   
   Second, increasing $V_{tip}$ favors the tunneling of electrons through the barrier as it is being reduced. The static potential becomes steeper in the vicinity of the tip, thus the probability an electron can tunnel after absorbing only one or two photons can become dominant. In other words, photofield emission can overcome multiphoton absorption. Interesting results have been obtained and extensively studied in this regime by Yanagisawa {\it et al.} \cite{Yanagisawa2011}.

\begin{figure}
\begin{center}
\begin{minipage}{100mm}
\includegraphics[width = 100 mm]{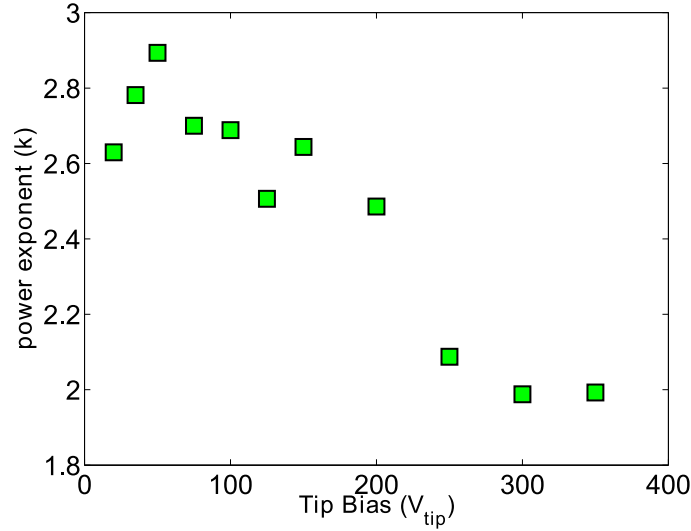}%
\caption{\label{fig:exponent}Power law exponent as a function of tip bias ($V_{tip}$). This exponent $k$ is retrieved from fit of the power dependence of the form $\alpha I^{k}$.}
\end{minipage}
\end{center}
\end{figure}

To determine the power law, for each $V_{tip}$, the number of electrons versus the laser intensity $I$ is fitted with a polynomial function of the form $\alpha I^{k}$, where $k$ can be non-integer. Indeed, in the mean-field approach, including photofield emission and ATP, if the system absorbs $n$ photons before an electron is emitted, we expect the exponent, $k$, to be equal to $n$. The energy distribution of the electrons will average to $k$ over all the contributions of the different number of absorbed photons. This exponent $k$ is plotted in Figure~\ref{fig:exponent} as a function of $V_{tip}$. Here the laser intensity is scanned from 2.5 to 7.5 x10$^{11}$~W/cm$^{2}$ and the exponent is retrieved from a polynomial fit of the data. The continuously decreasing behavior of this curve is similar to that observed by Barwick {\it et al.} \cite{barwick_laser-induced_2007}, although shifted by an offset which can be attributed to different experimental parameters. If the vacuum is not sufficient, or if the tip is not properly cleaned, surface contamination of the tip can alter the metallic work function. These retrieved exponents corroborate with the polarization curve as seen in Figure~\ref{fig:setup}(b), which was taken for $V_{tip} =$ 50~V and fits well with an exponent of 2.8.

\begin{figure}
\begin{center}
\begin{minipage}{100mm}
\includegraphics[width = 100mm]{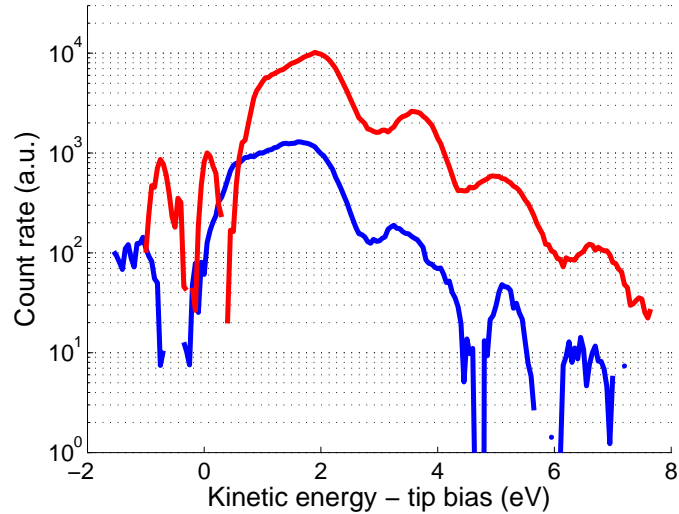}%
\caption{\label{fig:spectra}(color online) Two photoelectron energy spectra for a laser with mean power of 35~mW (2x10$^{11}$~W/cm$^{2}$) (blue) and 70~mW (4x10$^{11}$~W/cm$^{2}$) (red). In this case, the applied bias is 20~V. Distinct peaks are visible--separated exactly by the photon energy--a clear signature of above-threshold photoemission. Low tip bias allows the maximization in contrast of the peaks, also partially due to the relative resolution of the spectrometer. These spectra are obtained with a resolution of 0.05~V on the spectrometer and with a Savitzky-Golay smoothing filter over 16 adjacent points. The contrast is about 45~\%, defined as between the maximum of each peak and the average of the two adjacent minima. This contrast is extremely sensitive to experimental parameters and can vary from tip to tip.}%
\end{minipage}
\end{center}
\end{figure}

In a second set of experiments, details of the laser-induced emission regime are studied via energy resolved measurements of the electrons using the field retarding spectrometer.  In this case, the applied bias is 20~V leading to a value of the static electric field of $\sim$4$\times$10$^7$ V/m. For such a low tip bias, no thermal effects are expected. The investigations of thermally enhanced field emission performed by Kealhofer \textit{et al} \cite{Kealhofer2012} demonstrate that for similar laser power, emission is mainly due to multiphoton absorption in HfC tips, until the static electric field becomes on the order of 1 GV/m. Only above this value does thermally enhanced field emission become dominant. Figure~\ref{fig:spectra} presents two photoelectron energy spectra for a laser intensity of 1.8x10$^{11}$~W/cm$^{2}$ (blue) and 3.6x10$^{11}$~W/cm$^{2}$ (red). These spectra are obtained with a scan resolution of 0.05~V over 500~ms and with a Savitzky-Golay smoothing filter over 16 adjacent points. Distinct peaks separated by the photon energy are visible, which are a clear signature of ATP, similar to what has been observed by Kr\"uger {\it et al.} \cite{kruger_attosecond_2011}. For the high-energy spectrum, we reach for the first three peaks a high constrast of about 45~\%. This contrast is defined between the maximum of each peak and the average of the two adjacent minima, and is extremely sensitive to experimental parameters and can vary from tip to tip. A small tip radius maximizes the field enhancement allowing the use of a low $V_{tip}$ (below 30 V) which enables the use of the spectrometer in its region of highest resolution. This maximizes the contrast of the peaks. 

To confirm these emission mechanisms, an energy-selective, laser-power dependence measurement of these ATP peaks was performed. Results are shown in Figure~\ref{fig:power} for an experiment with $V_{tip} = 30$~V and laser peak intensities from 1 to 3.6x10$^{11}$~W/cm$^{2}$, individual peaks are indicated by the same color throughout the figure. Figure~\ref{fig:power}(a) shows the energy spectrum recorded at 3.6$\times$10$^{11}$~W/cm$^{2}$, where different ATP peaks are identified with different colors. Figure~\ref{fig:power}(b) shows the electron count rate versus laser intensity plotted in log scale. Each line corresponds to one peak of the photoelectron spectra as defined in Figure~\ref{fig:power}(a).  The electron yield is obtained by integrating over all of the energy window corresponding to each peak. We retrieve the exponent of each energy peak by extracting the slope of each fitted line. Figure~\ref{fig:power}(c) shows the retrieved power exponents of the four peaks from about 3 to 6, with error bars relative to the fit. 

As expected for ATP, each peak corresponds to a different number of absorbed photons, therefore corresponding to a different exponent. The first peak is close to 3 (3.4 $\pm$ 0.1), since the work function of tungsten is roughly equal to 3 times the photon energy. The second, third and fourth peaks have respective exponents of {.0 $\pm$ 0.1, 5.0 $\pm$ 0.1, and 6.5 $\pm$ 0.3 showing that in this part of the spectrum, the system absorbs 4, 5 and 6 photons before electrons leave the laser focus. This confirms that in our case, the photoelectron emission is mainly due to multiphoton absorption, without any contribution of photofield emission. These experiments were reproduced for the two first peaks with various experimental conditions (different tips, different focusing), and with different energy selections--choosing electrons from within the entire peak, or just at the maximum of the peak. The values of the exponents are reproducible to within 10~\% for several tips.

\begin{figure}
\begin{center}
\begin{minipage}{150mm}
\subfigure[]{
\resizebox*{5cm}{!}{\includegraphics{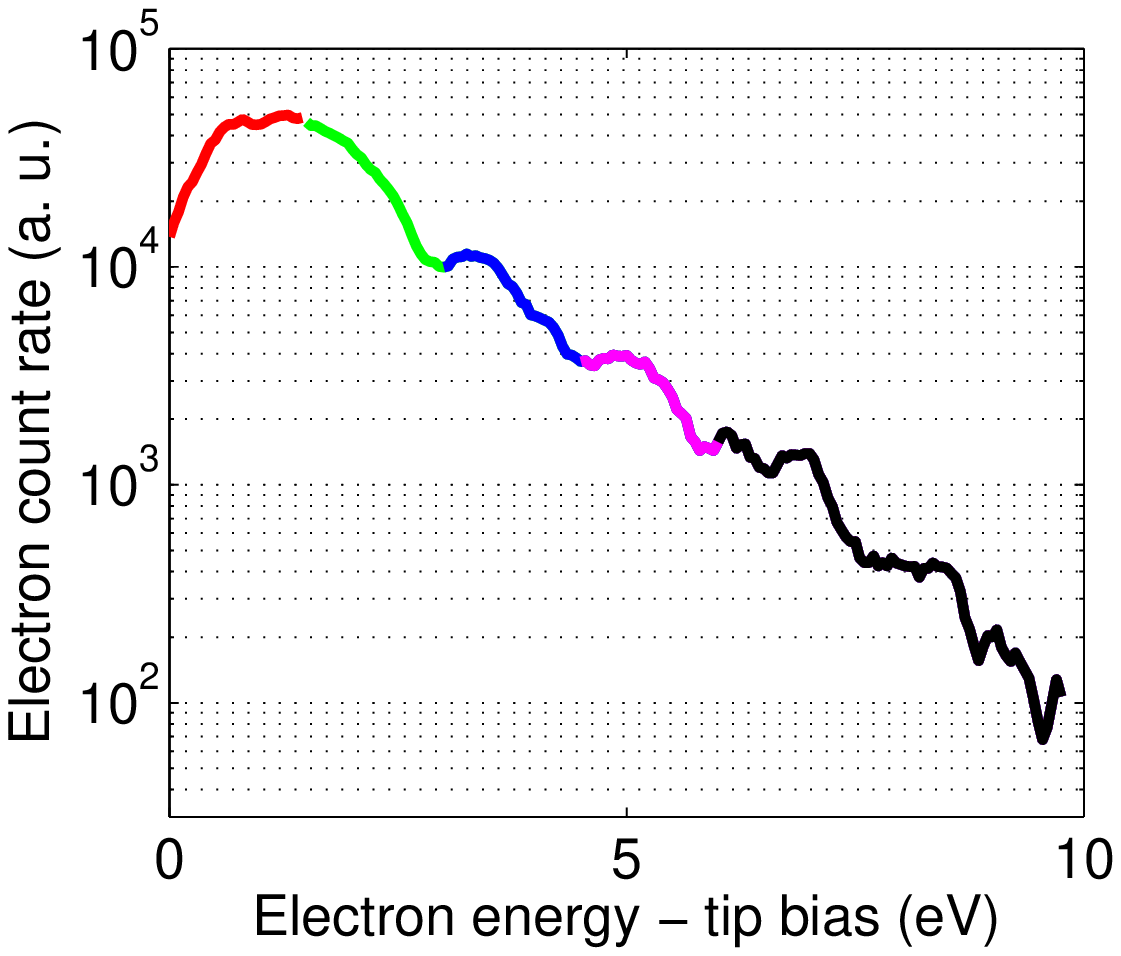}}}%
\subfigure[]{
\resizebox*{5cm}{!}{\includegraphics{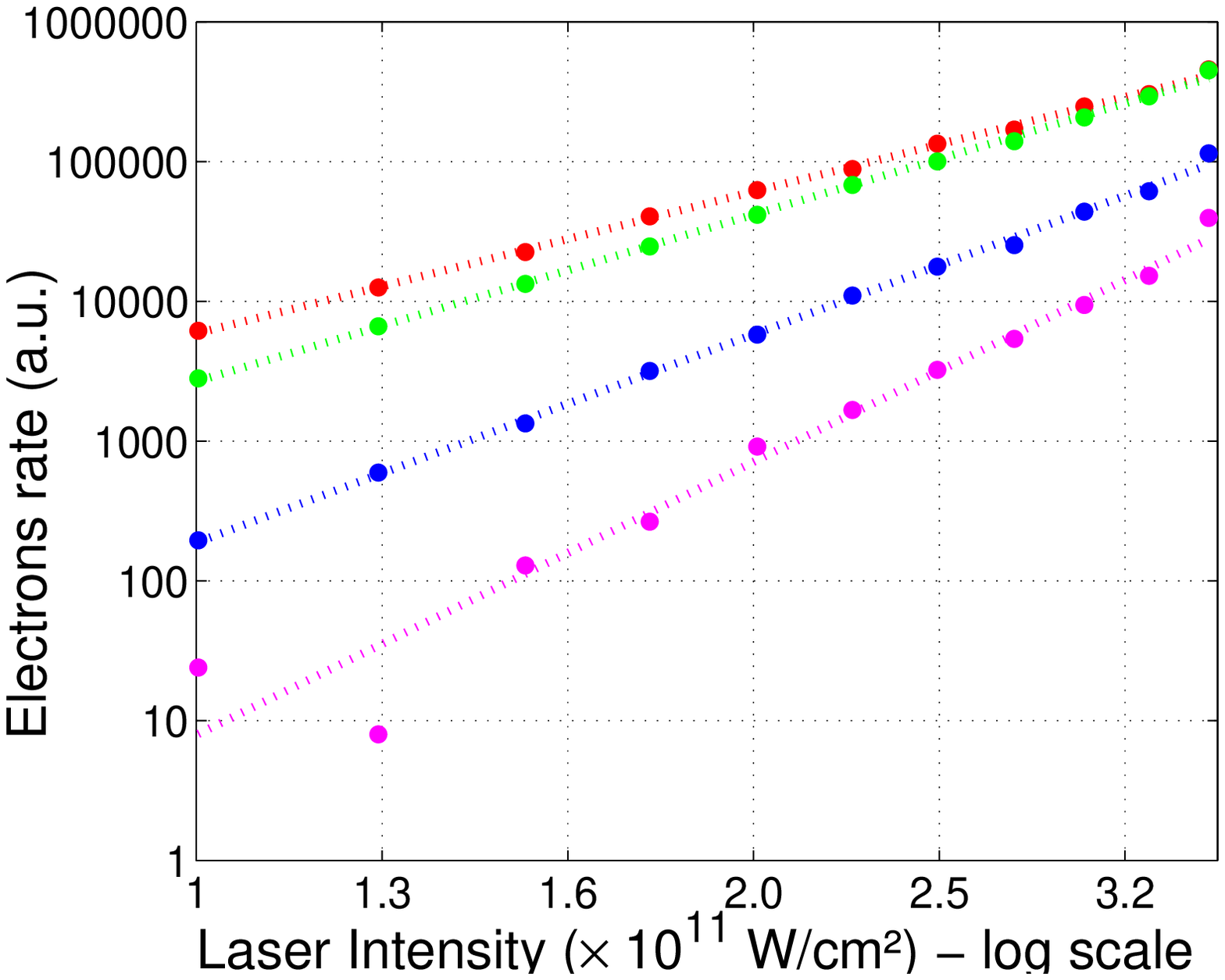}}}%
\subfigure[]{
\resizebox*{5cm}{!}{\includegraphics{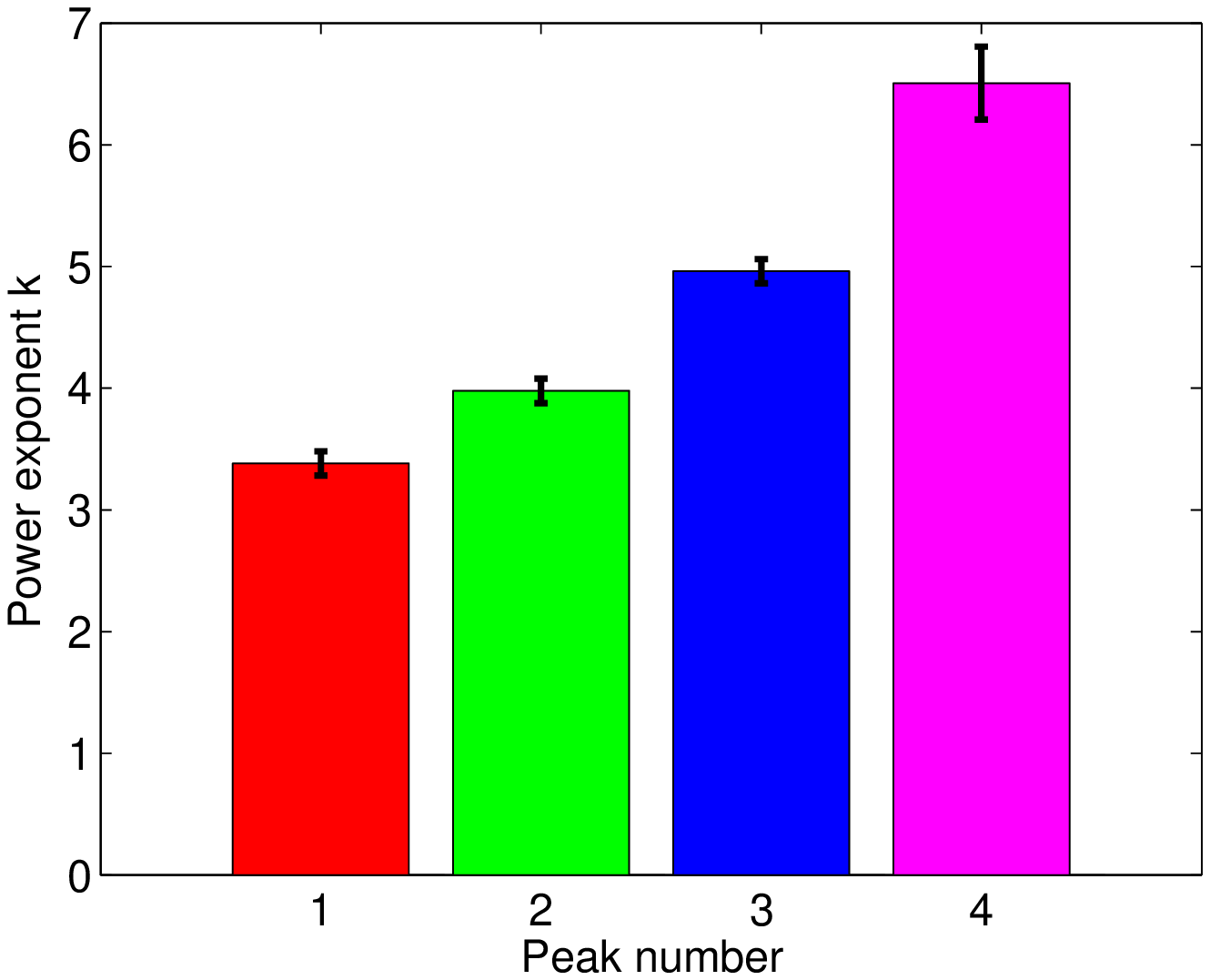}}}%
\begin{center}
\begin{minipage}{100mm}
\caption{\label{fig:power}(color online) (a) Electron energy spectrum obtained for a $V_{tip} = 30$~V and laser intensity of 3.6x10$^{11}$~W/cm$^{2}$. The different ATP peaks are identified by the different colors. (b) Electron yield versus laser intensity plotted for each ATP peak. The dots are the experimental values, and the lines are the fit from which we retrieve the exponent. (c) Power exponents for each peak. As expected for ATP, we obtain exponents from 3 to 6.}%
\end{minipage}
\end{center}
\end{minipage}
\end{center}
\end{figure}

\section{Conclusions}

In this paper we demonstrated the possibility to consistently reach the multiphoton regime for laser-induced electron emission. A clear above threshold photoemission signature has been observed with highly contrasted photoelectron spectra. A spectrally resolved power law study has been successfully implemented confirming a well-defined multiphotonic regime dominant over other mechanisms of emission. These studies open the route to control experiment in strong field using the light confinement induced by a single nanotip, extending the experiments performed on atoms and molecules.

\section{Acknowledgements}
Support by Marie Curie ITN ``FASTQUAST", ANR contract labex NEXT N° 11 LABEX 075 and by the region Midi-Pyr\'en\'ees is gratefully acknowledged. We thank P. Hommelhoff, H. Batelaan, J. Vigu\'e, P. Labastie and C. Meier for fruitful discussions, E. Baynard, D. Castex and M. Gianesin, for technical supports. L. Polizzi is particularly acknowledged for the strong technical support in designing the vacuum apparatus.  

\bibliographystyle{tMOP}
\bibliography{JMO_v4}

\begin{thebibliography}{27}
\providecommand{\natexlab}[1]{#1}

\bibitem[1]{rabitz_whither_2000}
Rabitz, H.; Vivie-Riedle, R.d.; Motzkus, M.; et~al. Whither the Future of
  Controlling Quantum Phenomena?.  {\em Science}  {\bf 2000}, {\em 288} (5467),
  824--828.  {PMID:} 10796997.

\bibitem[2]{meshulach_coherent_1998}
Meshulach, D.; Silberberg, Y. Coherent quantum control of two-photon
  transitions by a femtosecond laser pulse.  {\em Nature}  {\bf 1998}, {\em
  396} (6708), 239--242.

\bibitem[3]{monmayrant_newcomers_2010}
Monmayrant, A.; Weber, S.; Chatel, B. A newcomer's guide to ultrashort pulse
  shaping and characterization.  {\em Journal of Physics B: Atomic, Molecular
  and Optical Physics}  {\bf 2010}, {\em 43} (10), 103001.

\bibitem[4]{kupersztych_ponderomotive_2001}
Kupersztych, J.; Monchicourt, P.; Raynaud, M. Ponderomotive Acceleration of
  Photoelectrons in Surface-Plasmon-Assisted Multiphoton Photoelectric
  Emission.  {\em Physical Review Letters}  {\bf 2001}, {\em 86} (22),
  5180--5183.

\bibitem[5]{zherebtsov_controlled_2011}
Zherebtsov, S.; Fennel, T.; Plenge, J.; Antonsson, E.; Znakovskaya, I.; Wirth,
  A.; Herrwerth, O.; S\"ussmann, F.; Peltz, C.; Ahmad, I.; Trushin, S.A.;
  Pervak, V.; Karsch, S.; Vrakking, M.J.J.; Langer, B.; Graf, C.; Stockman,
  M.I.; Krausz, F.; R\"uhl, E.; et~al. Controlled near-field enhanced electron
  acceleration from dielectric nanospheres with intense few-cycle laser fields.
   {\em Nature Physics}  {\bf 2011}, {\em 7} (8), 656--662.

\bibitem[6]{PhysRevLett.108.237602}
Schertz, F.; Schmelzeisen, M.; Kreiter, M.; Elmers, H.J.; et~al. Field Emission
  of Electrons Generated by the Near Field of Strongly Coupled Plasmons.  {\em
  Phys. Rev. Lett.}  {\bf 2012}, {\em 108}, 237602.

\bibitem[7]{Kruger2012}
Kr{\"u}ger, M.; Schenk, M.; F{\"o}rster, M.; et~al. Attosecond physics in
  photoemission from a metal nanotip.  {\em Journal of Physics B: Atomic,
  Molecular and Optical Physics}  {\bf 2012}, {\em 45} (7), 074006.

\bibitem[8]{barwick_laser-induced_2007}
Barwick, B.; Corder, C.; Strohaber, J.; Chandler-Smith, N.; Uiterwaal, C.;
  et~al. Laser-induced ultrafast electron emission from a field emission tip.
  {\em New Journal of Physics}  {\bf 2007}, {\em 9} (5), 142--142.

\bibitem[9]{ropers_localized_2007}
Ropers, C.; Solli, D.R.; Schulz, C.P.; Lienau, C.; et~al. Localized Multiphoton
  Emission of Femtosecond Electron Pulses from Metal Nanotips.  {\em Physical
  Review Letters}  {\bf 2007}, {\em 98} (4), 043907.

\bibitem[10]{schenk_strong-field_2010}
Schenk, M.; Kr\"uger, M.; Hommelhoff, P. Strong-Field Above-Threshold
  Photoemission from Sharp Metal Tips.  {\em Physical Review Letters}  {\bf
  2010}, {\em 105} (25), 257601.

\bibitem[11]{kruger_attosecond_2011}
Kr\"uger, M.; Schenk, M.; Hommelhoff, P. Attosecond control of electrons
  emitted from a nanoscale metal tip.  {\em Nature}  {\bf 2011}, {\em 475}
  (7354), 78--81.

\bibitem[12]{park_strong_2012}
Park, D.J.; Piglosiewicz, B.; Schmidt, S.; Kollmann, H.; Mascheck, M.; et~al.
  Strong Field Acceleration and Steering of Ultrafast Electron Pulses from a
  Sharp Metallic Nanotip.  {\em Physical Review Letters}  {\bf 2012}, {\em 109}
  (24), 244803.

\bibitem[13]{herink_field-driven_2012}
Herink, G.; Solli, D.R.; Gulde, M.; et~al. Field-driven photoemission from
  nanostructures quenches the quiver motion.  {\em Nature}  {\bf 2012}, {\em
  483} (7388), 190--193.

\bibitem[14]{Fowler-nordheim}
{Fowler}, R.H.; {Nordheim}, L. {Electron Emission in Intense Electric Fields}.
  {\em Royal Society of London Proceedings Series A}  {\bf 1928}, {\em 119},
  173--181.

\bibitem[15]{gomer_field_1961}
Gomer, R. Field emission and field ionization. In: ;  Vol. 34,   : , 1961.

\bibitem[16]{keldysh_ionization_1965}
Keldysh, L. Ionization in a field of a strong electromagnetic wave.  {\em
  Soviet Physics Jetp-Ussr}  {\bf 1965}, {\em 20} (5), 1307.

\bibitem[17]{hommelhoff_ultrafast_2006}
Hommelhoff, P.; Kealhofer, C.; Kasevich, M.A. Ultrafast Electron Pulses from a
  Tungsten Tip Triggered by Low-Power Femtosecond Laser Pulses.  {\em Physical
  Review Letters}  {\bf 2006}, {\em 97} (24), 247402.

\bibitem[18]{Yanagisawa2011}
Yanagisawa, H.; Hengsberger, M.; Leuenberger, D.; Kl\"ockner, M.; Hafner, C.;
  Greber, T.; et~al. Energy Distribution Curves of Ultrafast Laser-Induced
  Field Emission and Their Implications for Electron Dynamics.  {\em Phys. Rev.
  Lett.}  {\bf 2011}, {\em 107} (Aug), 087601.

\bibitem[19]{agostini_free-free_1979}
Agostini, P.; Fabre, F.; Mainfray, G.; Petite, G.; et~al. Free-Free Transitions
  Following Six-Photon Ionization of Xenon Atoms.  {\em Physical Review
  Letters}  {\bf 1979}, {\em 42} (17), 1127--1130.

\bibitem[20]{PhysRevLett.59.1092}
Freeman, R.R.; Bucksbaum, P.H.; Milchberg, H.; Darack, S.; Schumacher, D.;
  et~al. Above-threshold ionization with subpicosecond laser pulses.  {\em
  Phys. Rev. Lett.}  {\bf 1987}, {\em 59}, 1092--1095.

\bibitem[21]{Mainfray1991}
Mainfray, G.; Manus, G. Multiphoton ionization of atoms.  {\em Reports on
  Progress in Physics}  {\bf 1991}, {\em 54} (10), 1333.

\bibitem[22]{blaga_strong-field_2009}
Blaga, C.I.; Catoire, F.; Colosimo, P.; Paulus, G.G.; Muller, H.G.; Agostini,
  P.; et~al. Strong-field photoionization revisited.  {\em Nature Physics}
  {\bf 2009}, {\em 5} (5), 335--338.

\bibitem[23]{banfi_experimental_2005}
Banfi, F.; Giannetti, C.; Ferrini, G.; Galimberti, G.; Pagliara, S.; Fausti,
  D.; et~al. Experimental Evidence of Above-Threshold Photoemission in Solids.
  {\em Physical Review Letters}  {\bf 2005}, {\em 94} (3), 037601.

\bibitem[24]{Wachter_Burgdorfer_PRB2012}
Wachter, G.; Lemell, C.; Burgd\"orfer, J.; Schenk, M.; Kr\"uger, M.; et~al.
  Electron rescattering at metal nanotips induced by ultrashort laser pulses.
  {\em Phys. Rev. B}  {\bf 2012}, {\em 86}, 035402.

\bibitem[25]{iaconis_spectral_1998}
Iaconis, C.; Walmsley, I. Spectral phase interferometry for direct
  electric-field reconstruction of ultrashort optical pulses.  {\em Optics
  Letters}  {\bf 1998}, {\em 23} (10), 792--794.

\bibitem[26]{Hommelhoff2006}
Hommelhoff, P.; Sortais, Y.; Aghajani-Talesh, A.; et~al. Field Emission Tip as
  a Nanometer Source of Free Electron Femtosecond Pulses.  {\em Phys. Rev.
  Lett.}  {\bf 2006}, {\em 96} (Feb), 077401.

\bibitem[27]{Kealhofer2012}
Kealhofer, C.; Foreman, S.M.; Gerlich, S.; et~al. Ultrafast laser-triggered
  emission from hafnium carbide tips.  {\em Physical Review B}  {\bf 2012},
  {\em 86} (3), 035405.

\end{thebibliography}

\label{lastpage}

\end{document}